\documentstyle[eqsecnum,multicol,aps,prl,epsf]{revtex}

\draft
\tighten
\begin{document}
\def\sqr#1#2{{\vcenter{\hrule height.3pt
      \hbox{\vrule width.3pt height#2pt  \kern#1pt
         \vrule width.3pt}  \hrule height.3pt}}}
\def\square{\mathchoice{\sqr67\,}{\sqr67\,}\sqr{3}{3.5}\sqr{3}{3.5}}
\def\today{\ifcase\month\or
  January\or February\or March\or April\or May\or June\or July\or
  August\or September\or October\or November\or December\fi
  \space\number\day, \number\year}

\def\Bbb{\bf}

\newcommand {\be}{\begin{equation}}
\newcommand {\ee}{\end{equation}}
\newcommand {\bea}{\begin{array}}
\newcommand {\cl}{\centerline}
\newcommand {\eea}{\end{array}}
\newcommand{\bn}{\begin{eqnarray}}
\newcommand{\en}{\end{eqnarray}}
\renewcommand {\thefootnote}{\fnsymbol{footnote}}

\def\nc{noncommutative }
\def\com{commutative }
\def\ncy{noncommutativity }
\def\repr{representation }
\def\Ham{Hamiltonian }
\def\reps{representations }
\def \simlt{\stackrel{<}{{}_\sim}}
\def \simgt{\stackrel{>}{{}_\sim}}
%\def{\th}{\theta}
%\baselineskip 0.65 cm

%%%%%%%%%%%%%%%%%%%%%%%%%%%%%%%%%%%%%%%%%%%%%%%%%%%%%%%%%%%%%%%%%
%%%% Beware of the kludge to put indices on the author names %%%%
%%%%%%%%%%%%%%%%%%%%%%%%%%%%%%%%%%%%%%%%%%%%%%%%%%%%%%%%%%%%%%%%%

\title{Comments on the Hydrogen Atom Spectrum in the Noncommutative 
Space}

\author{M. Chaichian, M.M. Sheikh-Jabbari$^{\dagger}$ and A. Tureanu}

\address {High Energy Physics Division, Department of Physical Sciences, University of Helsinki\\ and \\
Helsinki Institute of Physics, P. O. Box 64, FIN-00014 Helsinki, Finland \\
E-mail: Masud.Chaichian@helsinki.fi, atureanu@pcu.helsinki.fi \\
$^{\dagger}$ 
Department of Physics, Stanford University, Stanford, CA 94305-4060, USA\\
E-mail: jabbari@itp.stanford.edu}

\maketitle

%\centerline{\bf {$\frak{D}\frak{R}\frak{A}\frak{F} \frak {T}$}}

\begin{abstract}

There has been disagreement in the literature on whether the hydrogen atom
spectrum receives any tree-level correction due to noncommutativity. Here
we shall clarify the issue and show that indeed a general argument on
the structure of proton as a nonelementary particle leads to the
appearance of such corrections. As a showcase, we evaluate the corrections 
in a simple nonrelativistic quark model with a result in  agreement 
with the previous one we had obtained by considering the electron moving 
in the external electric field of proton. Thus the previously obtained 
bound on the noncommutativity parameter, $\theta < (10^4 GeV)^{-2}$, 
using the Lamb shift data, remains valid.

\end{abstract}

\pacs{PACS: 11.15.-q, 11.30.Er, 11.25.Sq.
{\qquad} {\qquad} $\ \ \ \ $
%{\qquad} {\qquad}{\qquad} {\qquad}
HIP-2002-67/TH
}
\vspace*{0.1cm}

\begin{multicols} {2}

%\tableofcontents

%\section{Introduction}
\setcounter{equation}{0}

Recently a large amount of research work has been devoted to the study of physics on 
\nc space-times and in particular \nc Moyal plane (for a review see, e.g., 
\cite{review}). In these works both quantum mechanics (QM) and field theory on 
\nc spaces have been studied. Besides the theoretical interests, by comparing the 
results of \nc version of usual physical models with present data, lower bounds 
on the \nc scale $\Lambda_{NC}$ have been obtained 
\cite{{Frank},{Harvey},{Roiban}}: as a conservative estimate, 
$\Lambda_{NC}\simgt 1-10\ TeV$. 

In this commentary we would like to focus on the Hydrogen atom in the \nc 
space and re-analyze its spectrum. This system has previously been considered 
in \cite{{Lamb},{HK}} with a disagreement on the results. Here, through a more 
careful analysis we intend to clarify the discrepency. In \cite{Lamb} we 
have analyzed the Hydrogen atom in the \nc space considering the system 
described by a one-particle Schrodinger equation. Explicitly we considered the 
electron 
in an external Coulomb field; hence the system is described by
\[
H|\psi\rangle=i{\hbar} {\partial\over \partial t} |\psi\rangle 
\]
with 
\be 
H={\hat{p}_e^2\over 2m_e}+V(\hat{x}_e)\ ,
\ee
where $\hat{x}_e$ and $\hat{p}_e$ are the phase space coordinates of the 
electron and 
\bn\label{NCXP}
\left [\hat{x}_e^i,\hat{x}_e^j \right ]&=&i\theta^{ij}\ , \cr
\left [\hat{x}_e^i,\hat{p}_e^j\right ]&=&i{\hbar}\delta^{ij}\ , \cr
\left [\hat{p}_e^i,\hat{p}_e^j\right ]&=&0\ .
\en
Then it is easy to see that the new coordinates
$x_i=\hat{x}_i+\frac{1}{2\hbar}\theta_{ij}\hat{p}_j\ ,\;
p_i=\hat{p}_i\ $ satisfy the usual canonical commutation
relations \cite{Lamb}
\bn
\left [\hat{x}_e^i,\hat{x}_e^j \right ]&=&0\ , \cr
\left [\hat{x}_e^i,\hat{p}_e^j\right ]&=&i{\hbar}\delta^{ij}\ , \cr
\left [\hat{p}_e^i,\hat{p}_e^j\right ]&=&0\ ,
\en
and in terms of these ``canonical'' coordinates, the 
Hamiltonian takes the familiar form of usual Hydrogen atom plus \nc corrections
({\it cf.} Eq.(2.5) in Ref.\cite{Lamb}),

\begin{eqnarray}\label{NCppp}
V(r,p)=
%\frac{Ze^{2}}{\sqrt{(x_i-\theta_{ij}p_j/2\hbar)(x_i-\theta_{ik}p_k/2\hbar)}}\cr
%&=&-\frac{Ze^{2}}{r}-Ze^{2}\frac{x_i\theta_{ij}p_j}{2\hbar
%r^{3}}+ O(\theta^2) \cr &=&
-\frac{Ze^{2}}{r}-Ze^{2}\frac{\vec{L}\cdot\vec{\theta}}{4\hbar r^{3}}+ O(\theta^2)\ ,
\end{eqnarray}
where $\theta_{i}=\epsilon_{ijk}\theta_{jk}$, $r=\sqrt{\sum_ix_i^2}$ and $\vec{L}=\vec{r}\times \vec{p}$.
(The value of $|\vec{\theta}|$ is the inverse square of the \nc scale 
$\Lambda_{NC}$.)
From here we concluded that there exist \nc corrections to the spectrum and 
comparing our results with the data for the Lamb-shift experiments, we obtained 
the bound $\Lambda_{NC} \simgt 10\ TeV$.
 
On the other hand, in a more detailed analysis, the nucleus (here the proton) 
which exerts the Coulomb potential should also be considered as a dynamical 
object. In other words, one should solve the two-body Schrodinger 
equation. 
For the \nc Hydrogen atom this has been done in  Ref.\cite{HK}. 
There it was assumed that the proton, similar to the electron, is described by 
NC QED \cite{{Haya},{Ihab}}. Based on this assumption and the fact that under 
charge conjugation $\theta_{ij}$ changes the sign \cite{CPT}, it was shown that the \ncy effects will 
not change the spectrum of this two-body problem at the tree level ({\it cf.}
Eq.(28) of Ref.\cite{HK}), in disagreement with the results of 
\cite{Lamb} (in particular, with the above Eq. (\ref{NCppp})).

In this note we argue that in fact the assumption of  
Ref.\cite{HK} that the effective \ncy parameter for proton is equal to that of 
electron with a minus sign is not physically valid. Therefore, the 
``cancellation'' of \ncy effects is not complete and hence our previous results 
on the form (\ref{NCppp}) for the potential with the correction term, as well as on the lower bound on $\Lambda_{NC}$ are indeed valid.
The essential point is that the proton, due to the fact that it has structure and is a composite particle, cannot be described by NC QED (applicable to elementary
particles). To systems such as positronium, however, the analysis of Ref. \cite{HK} is applicable, resulting in no corrections to the
spectrum at the tree level, due to \ncy of space-time. Noting the conservative bounds on 
$\Lambda_{NC}\simgt 1-10\ TeV$ obtained from other physical analysis 
\cite{{Frank},{Harvey}}, and that $\Lambda_{QCD}$, or the inverse of the proton size, is of the order of $200\ MeV$, we notice that
${\Lambda_{QCD} \over\Lambda_{NC}}\ll 1$. In other words the QCD effects 
(here the internal structure of proton)  become important much before the 
\nc effects. In short, proton in the noncommutative Hydrogen atom 
essentially behaves as a {\it commutative} particle.

A full analysis of the problem, however, needs a better understanding of the 
noncommutative Standard Model (NCSM). 
%i.e. the electromagnetic couplings to particles with charges other than 
%$\pm1,0$ (the charge quantization problem). 
Unfortunately, despite several 
efforts 
in constructing such a model \cite{{NCSM},{Wess}}, a complete formulation of 
NCSM is not yet available. Therefore, to present a quantitave treatment of the issue with the 
above arguments, we try to takle the problem of finding an effective 
description of electromagnetic interactions of proton through a naive
quark model.
In such a model we can safely assume
 that inside the proton we deal with free quarks. 

However, there still remains a major problem to be addressed: in the 
NCQED the only possible charges coupled to photon are $\pm 1, 0$ \cite{{Haya},{nogo}}. 
As a result quarks (with
fractional charges) cannot be described by NCQED.
Nevertheless, since we are interested only in first order effects in $\theta$, 
we can use NCQED vertices for quarks, 
\footnote{One may also try to use the expressions given in 
Refs.\cite{{NCSM},{Wess}} for the vertices. In that case, although the 
numerical
coefficients (for the second term in (\ref{NCppp}), as derived from (\ref{pot}) and (\ref{Vq})) would be slightly 
different from what we present here, the order of the bound on 
$\Lambda_{NC}$ obtained in this way would be the same.} though only up to 
first order in $\theta$. 
Then the effective electron-proton interaction is the sum of the 
electron-quark Coulomb potentials for $u$, $u$ and $d$ quarks, namely,
\be\label{pot}
V=V_{u_1}+V_{u_2}+V_{d}\ .
\ee
The expressions for the potentials $V_q$ can be obtained, as usual, as 
the nonrelativistic limit of \nc one-photon exchange,
\be\label{Vq}
V_q=-Qe^2 V(\vec{x}_q+{1\over 4}\vec{\theta}\times \vec{K}_q)\ ,
\ee
where $V(\vec{r})={1\over |\vec{r}|}$, $Q$ is $2/3$ and $-1/3$ for $u$ and $d$ 
quarks respectively, $\vec{x}_q$ is the relative separation of electron 
and the corresponding quark and $\vec{K}_q$ is $\vec{P}_e+\vec{P}_u$ for 
$u$ quark and $\vec{P}_e-\vec{P}_d$ for $d$ quark ($\vec{P}_u$ 
and $\vec{P}_d$ are the momenta of the corresponding quarks). The 
expression (\ref{Vq}) for the values of $Q=+1$ and
$\vec{K}_q=\vec{K}=\vec{P}_e+\vec{P}_{p}$, with $\vec{P}_{p}$ the 
momentum of 
proton, formally coincides with Eq. (28) of Ref. \cite{HK}.
However, in Ref.\cite{HK} this was used for 
the {\it overall} electron-proton potential.

We should emphasize that the expressions (\ref{Vq}) are valid up to 
the first order in $\theta$. 
%If we denote the fraction of proton momentum 
%carried by a single quark by $\xi_q$, i.e.
%$\vec{P}_u=\xi_u\vec{P}_p$, $\vec{P}_d=\xi_d\vec{P}_p$, then 
%$2\xi_u+\xi_d\lesssim 1$. 
Expanding (\ref{Vq}) to the first order in $\theta$, it is evident that 
the effective Coulomb potential of proton and electron does not only depend on 
the Hydrogen atom center of mass momentum, which is the sum of  electron and 
proton momenta, 
$\vec{P}_e +\vec{P}_p=\vec{P}_e +\vec{P}_{u_1}+
\vec{P}_{u_2}+\vec{P}_d$, 
%\footnote{
%Also note that one also needs to make an average over the quark positions, and 
%this will completely wash out the \ncy effects which may come from proton. 
%That is, proton behaves as a commutative particle.}
invalidating the result 
of  Ref.\cite{HK}.   
 
%As a by-product of similar rough quark model considerations one may 
%extract a lower bound on $\Lambda_{NC}$ using the data on neutron electric 
%dipole moment \cite{anupam}. 

As a by-product, based on a similar quark model consideration on the 
electric dipole moment of neutron, one obtains  the \ncy correction
$$
\vec{d}_n^{NC}\simeq -\sum_i|Q_i|\vec{\theta}\times\vec{P_{q_i}}
$$
with $|\vec{P}_u|\sim|\vec{P}_d|\sim\Lambda_{QCD}\sim 200\ MeV$ and 
therefore in this model $|\vec{d}_n^{NC}|\sim e {\Lambda_{QCD}\over 
\Lambda_{NC}^2}$. 
Using the experimental upper bound of $|\vec{d}_n|< 0.63\times 
10^{-25}\ e{\rm cm} $ \cite{Data},  one obtains the lower bound 
$\Lambda_{NC}\simgt 200\ TeV$.

The work of M.C. and A.T. is partially supported by the Academy of
Finland, under the Project No. 54023. 
The work of M.M. Sh.-J. is supported in part
by NSF grant PHYS-9870115 and in part by funds from the Stanford Institue for 
Theoretical Physics.

%%%%%%%%%%%%%%%%%%%%%%%%%%%%%%%%%%%%%%%%%%%%%%%%%%%%%%%%%%%%%%%%%%%%%%%%%%%%

\end{multicols}
\end{document}